\documentclass[11pt,twoside]{article}

\def \etal {et~al.~}
\def \Vobs {\ifmmode V_{\rm obs} \else $V_{\rm obs}$ \fi} 
\def \Vrot {\ifmmode V_{\rm rot} \else $V_{\rm rot}$ \fi} 
\def \Vvir {\ifmmode V_{\rm  vir} \else  $V_{\rm vir}$  \fi} 
\def \Vmax {\ifmmode V_{\rm  max} \else  $V_{\rm max}$  \fi} 
\def \kms  {\ifmmode  \,\rm km\,s^{-1} \else $\,\rm km\,s^{-1}  $ \fi }
\usepackage{asp2006}
\usepackage{epsf}
\usepackage{psfig}
\usepackage{lscape}
\usepackage{graphicx}

\begin{document}
\title{The Tully-Fisher Zero Point Problem}

\author{Aaron   A.    Dutton\altaffilmark{1},   Frank  C.    van   den
  Bosch\altaffilmark{2}, and St\'ephane Courteau\altaffilmark{3}}

\altaffiltext{1}{UCO/Lick Observatory and  Department of Astronomy and
  Astrophysics, University of California, Santa Cruz, CA 95064}

\altaffiltext{2}{Max-Planck-Institut  f\"ur  Astronomie,  K\"onigstuhl
  17, 69117 Heidelberg, Germany}
 
\altaffiltext{3}{Department of  Physics, Queen's University, Kingston,
  ON K7L 3N6 Canada}

\begin{abstract} 
  A  long  standing problem  for  hierarchical  disk galaxy  formation
  models has been  the simultaneous matching of the  zero point of the
  Tully-Fisher relation  and the galaxy luminosity  function (LF).  We
  illustrate this problem for a  typical disk galaxy and discuss three
  solutions: low  stellar mass-to-light ratios, low  initial dark halo
  concentrations,  and no  halo contraction.   We speculate  that halo
  contraction may  be reversed through a combination  of mass ejection
  through  feedback and  angular  momentum exchange  brought about  by
  dynamical friction  between baryons and dark matter  during the disk
  formation process.
\end{abstract}

\vspace{-0.1cm}
\section{Introduction}
The  relation between  the rotation  velocity and  luminosity  of disk
galaxies, commonly  known as the Tully-Fisher (TF)  relation (Tully \&
Fisher  1977), is  one  of  the most  fundamental  properties of  disk
galaxies.  This  is because it is  a link between  luminous mass (i.e.
baryons) and dynamical  mass (baryons and dark matter)  and because it
has    very   little    intrinsic   scatter    $\sigma_{\ln   V}\simeq
0.1$. Furthermore,  unlike the  Faber-Jackson relation for  early type
galaxies,  the scatter  in the  TF  relation is  {\it independent}  of
surface brightness, size, or offsets from the size-luminosity relation
(Courteau \& Rix 1999; Courteau \etal 2007).

CDM  based disk  galaxy formation  models  are able  to reproduce  the
slope, and  in some cases the  amount of scatter, in  the TF relation.
However, a  long standing  problem has been  the matching of  the zero
point  of  the TF  relation,  with  the  generic problem  of  galaxies
rotating  too fast  at  a given  luminosity.   This has  been seen  in
analytic  models (van den  Bosch 2000;  Mo \&  Mao 2000;  Dutton \etal
2007),   semi-analytic  models   (e.g.   Benson   \etal  2003   )  and
cosmological  N-body  simulations (e.g.   Eke,  Navarro, \&  Steinmetz
2001)

\section{The TF-LF Challenge}
The problem  of matching  the zero  point of the  TF relation  is made
worse  when additional  constraints,  such as  disk  sizes and  number
densities, are  placed on the  models.  Semi-analytic models  that are
able to simultaneously reproduce the TF zero point and LF require that
$\Vrot = V_{200}$ (e.g.   Somerville \& Primack 1999) or $\Vrot=\Vmax$
(e.g.  Croton  \etal 2006).   That is, these  do not account  for halo
contraction, or  the self-gravity  of the disk.   Here $\Vrot$  is the
observed  rotation  velocity  (corrected  for  inclination  and  other
observational effects), $V_{200}$ is the halo circular velocity at the
virial radius,  and $\Vmax$  is the maximum  circular velocity  of the
dark matter halo in the absence of dynamical evolution.  Bullock \etal
(2001)  showed that  $\Vmax\simeq  1.1-1.2 V_{200}$  for typical  halo
concentrations   in    a   $\Lambda$CDM   cosmology.     Support   for
$\Vrot\simeq\Vmax$  also comes  from more  direct  observational based
measurements of halo masses (Eke \etal 2006) and the velocity function
of isolated halos (Blanton \etal 2007).

Thus,  in  the  absence  of  baryonic effects,  there  appears  to  be
reasonable agreement  between theory and  observation.  Problems occur
when the contribution of the  baryons to rotation curve and the effect
of   halo  contraction   (Blumenthal  \etal   1986)  are   taken  into
account. These  effects increase  the observed rotation  velocity, and
typically result in $\Vrot \sim 2 V_{200}$ (e.g.  Navarro \& Steinmetz
2000; Dutton \etal 2007).

\begin{figure}
\begin{center}
\includegraphics[width=0.68\textwidth]{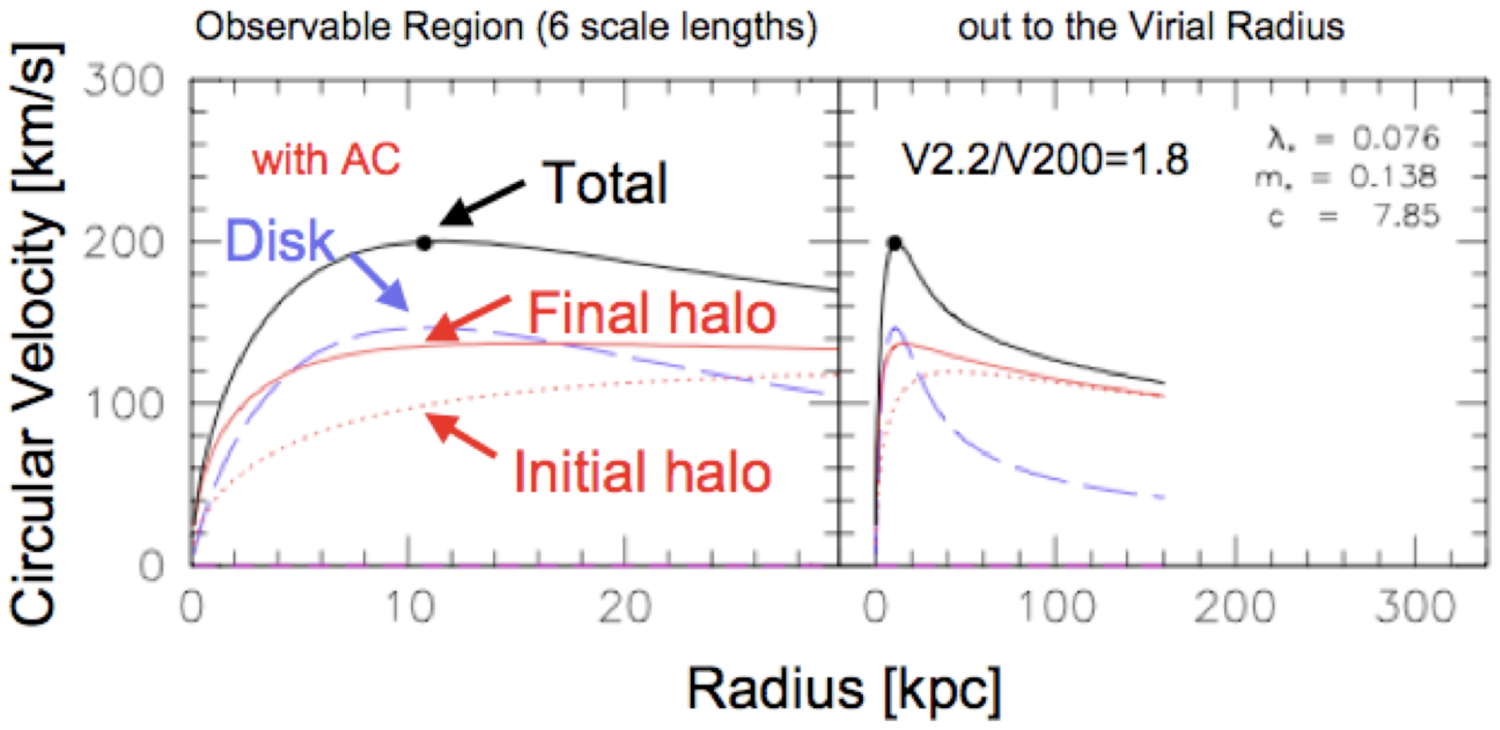}
\includegraphics[width=0.68\textwidth]{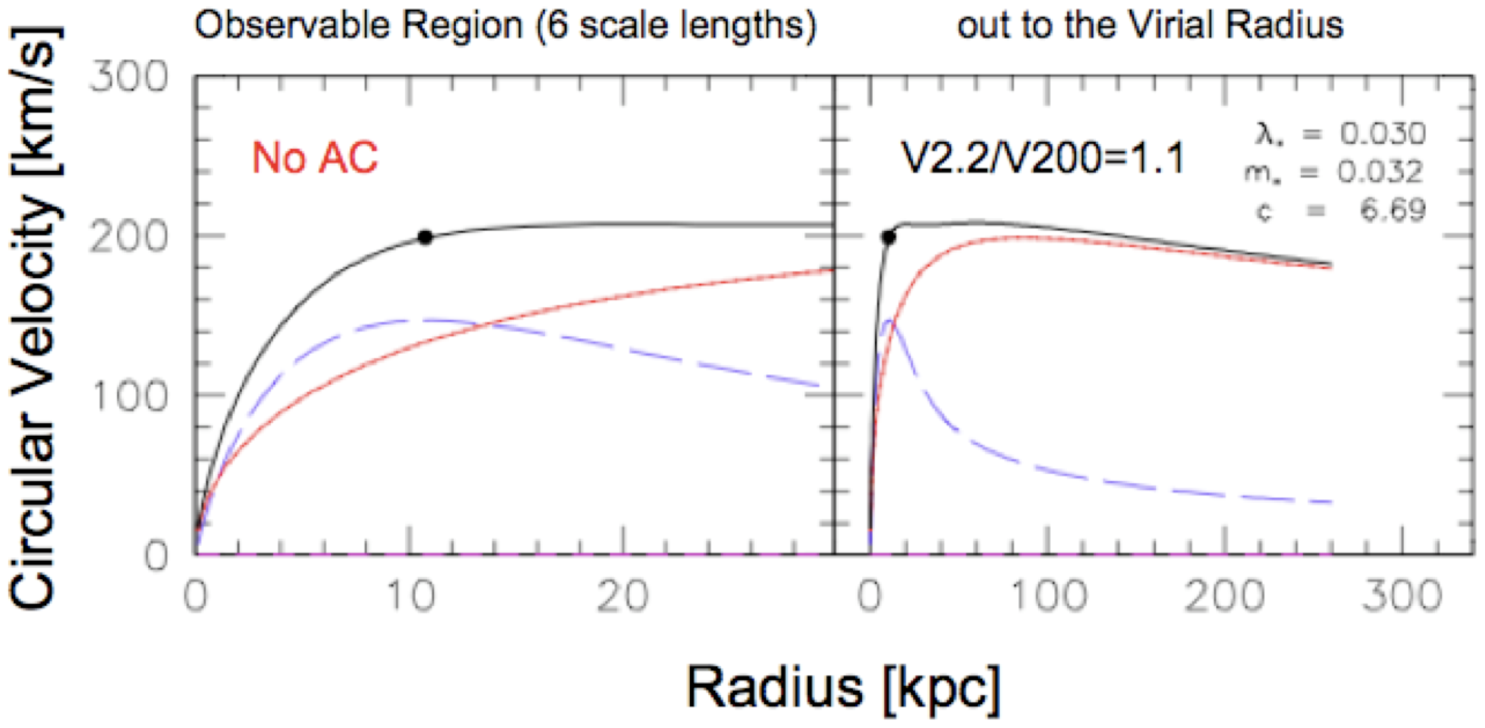}
\caption{Rotation  curves of model  galaxies with  $V_{2.2}=200 \kms$.
  These     models     are     constructed    to     reproduce     the
  velocity-luminosity-radius scaling relations as well as the relation
  between  concentration   and  virial  mass   in  $\Lambda$CDM.   The
  upper/lower  panels  show models  with/without  the  effect of  halo
  contraction.  The  halo contraction model requires  a low-mass halo,
  in  conflict  with  observations,  and  a high  spin  parameter,  in
  conflict  with   theoretical  expectations.   By   contrast  the  no
  contraction  model satisfies  the halo  mass constraint  and  uses a
  realistic galaxy spin parameter.}
\label{fig:VLRmu}
\end{center}
\end{figure}

\subsection{Illustration of the Problem}
As an illustration of the problem we consider the idealized case of an
exponential disk embedded in an  NFW halo. We pick $V_{2.2}=200 \kms$,
then from the TF relation (Courteau \etal 2007; Pizagno \etal 2007) we
know the typical I-band luminosity,  which is converted into a stellar
mass  using typical  galaxy colors  and  the relations  in Bell  \etal
(2003).   From  the  size-luminosity  relation  (Pizagno  \etal  2005;
Courteau \etal 2007)  we know the average scale  length of the galaxy.
From  cosmological  simulations  we  know  the  average  concentration
parameter of  dark matter halos of  a given mass  (Bullock \etal 2001;
Macci{\`o} \etal  2007). Here  we assume the  WMAP 3rd  year cosmology
($\sigma_8=0.78,      \Omega_0=0.268,      \Lambda=0.732,     \Omega_b=0.045, 
h=0.70$).  This model is fully specified, and we can solve for
the virial  mass (including the self  gravity of the  baryons and halo
contraction).  The resulting rotation  curve is shown in Fig.~1.  This
galaxy has  $V_{2.2}/V_{200} \sim 1.8$,  where $V_{2.2}$ is  the total
rotation velocity at 2.2R$_{\rm d}$.

\subsection{Possible solutions to the TF zero-point problem}
We now discuss the three possible solutions to this problem.

{\bf  Lowering the  stellar mass  to  light ratio:}  This reduces  the
contribution of  the disk to  $V_{2.2}$, which thus induces  less halo
contraction.  For our fiducial galaxy,  a $M_*/L$ reduction of 0.2 dex
is  required.  There is  no observationally  supported IMF  that would
achieve this,  and most dynamical  estimates of galaxy  masses suggest
baryons (disks  and bulges)  contribute substantially (i.e.   at least
half the mass) to $V_{2.2}$  (Courteau \& Rix 1999; Weiner \etal 2001;
Bershady \etal in these proceedings).

{\bf  Lowering the  initial  halo concentration:}  This preserves  the
inner structure  of the  halo while increasing  its virial  radius and
thus  reducing $V_{2.2}/V_{200}$.  However,  for $V_{2.2}/V_{200}=1.1$
an  initial concentration  of $\sim  3$  is required.   Such low  halo
concentrations require  cosmological parameters inconsistent  with the
WMAP 3rd year constraints (Spergel \etal 2007).

{\bf Turning off or reversing halo contraction}: Our model matches the
$V_{2.2}/V_{200}$ constraint  if we turn off halo  contraction.  A net
halo expansion  would be required if  we were to adopt  a smaller disk
scale length.

\section{Halo contraction}
For isolated  halos with smooth  cooling the halo contracts.   This is
well tested and well understood  (e.g.  Sellwood \& McGaugh 2005; Choi
\etal  2006).   However,   in  the  hierarchical  structure  formation
paradigm disk galaxies are not expected to form from smooth cooling in
isolated non-evolving halos.

Since cosmological simulations are still unable to make realistic disk
galaxies that follow the  scaling relations between rotation velocity,
luminosity  and  size  (see   Courteau  \etal  in  this  proceedings),
something must be missing.  Numerical resolution is certainly still an
issue, but more importantly, key physics relating to gas cooling, star
formation, and feedback is not yet understood.

\subsection{Is the reversal of halo contraction physically possible?}
Two mechanisms that are expected to occur during galaxy formation
have the ability to reverse the effects of halo contraction.
\begin{itemize}
\item {\bf Feedback:} If disks form adiabatically and a large fraction
  of  the disk mass  is removed  rapidly then  net halo  expansion can
  result (e.g.   Gnedin \&  Zhao 2002).  If  this process  is repeated
  several  times  a  substantial  reduction  in halo  density  can  be
  achieved (Read \& Gilmore 2005). Feedback also offers a mechanism to
  explain  why galaxy formation  is so  inefficient, since  cooling is
  very efficient in galaxy mass halos (e.g.  van den Bosch 2002).

\item  {\bf  Angular momentum  exchange:}  Dynamical friction  between
  baryons and dark  matter can cause the halo  to expand. This process
  can occur both in a formed disk through bars (Weinberg \& Katz 2002;
  Sellwood 2006),  or during the formation process  via large baryonic
  clumps (El Zant \etal 2001; Mo \& Mao 2004; Tonini \etal 2006).

\end{itemize}
A potential problem for  the angular momentum exchange solution occurs
if  the baryons  lose too  much angular  momentum, then  the resulting
disks will be too small. Furthermore, as baryons lose angular momentum
and move  to smaller  radii they  will drag in  the dark  matter halo,
which may cancel  out the halo expansion achieved  through the angular
momentum exchange.  Feedback can help solve these problems by ejecting
(low  angular momentum)  material  from the  centers  of galaxies  and
causing the halo to expand.

\acknowledgements %%% Text of acknowledgements runs on after this command.
AAD   acknowledges  financial  support   from  the   National  Science
Foundation  grant AST-0507483.   S.C.  acknowledges financial  support
from the National Science and Engineering Council of Canada.

\end{document}